\documentclass[aps,twocolumn,floatfix,groupedaddress,nofootinbib,showkeys]{revtex4-1}
\usepackage{csquotes}
\usepackage{graphicx,color}
\usepackage{float}
\usepackage[caption=false]{subfig}
\usepackage{multirow}
\usepackage{dsfont}

\begin{document}

\title{Evidence for high-velocity solid dust generation induced by runaway electron impact in FTU}
\author{M. De Angeli,$^{1}$ P. Tolias,$^{2}$  S. Ratynskaia,$^{2}$  D. Ripamonti,$^{3}$ L. Vignitchouk,$^{2}$ F. Causa,$^{1}$ G. Daminelli,$^{3}$ B. Esposito,$^{4}$ E. Fortuna-Zalesna,$^{5}$ F. Ghezzi,$^{1}$ L. Laguardia,$^{1}$ G. Maddaluno,$^{4}$ G. Riva$^{3}$ and W. Zielinski$^{5}$}
\address{$^1$Institute for Plasma Science and Technology - CNR, Via R. Cozzi 53, 20125 Milan, Italy\\
         $^2$Space and Plasma Physics - KTH Royal Institute of Technology - Teknikringen 31, 10044 Stockholm, Sweden\\
         $^3$Institute of Condensed Matter Chemistry and Energy Technologies - CNR, Via R. Cozzi 53, 20125 Milan, Italy\\
         $^4$ENEA, C.R. Frascati, Via E. Fermi 45, 00044 Frascati (Roma), Italy\\
         $^5$Warsaw University of Technology, 02-507 Warsaw, Poland}

\begin{abstract}
\noindent Post-mortem and in-situ evidence is presented in favor of the generation of high-velocity solid dust during the explosion-like interaction of runaway electrons with metallic plasma-facing components in FTU. The freshly-produced solid dust is the source of secondary de-localized wall damage through high-velocity impacts that lead to the formation of craters, which have been reproduced in dedicated light gas gun impact tests. This novel mechanism, of potential importance for ITER and DEMO, is further supported by surface analysis, multiple theoretical arguments and dust dynamics modelling.
\end{abstract}
\maketitle

\emph{Introduction}. $-$ The presence of dust in tokamaks constitutes an important issue with multi-faceted safety and operational implications for future fusion reactors\,\cite{dustint1,dustint2,dustint3}. In metallic armour tokamaks, the main solid dust generation mechanisms are the flaking of loosely-bound co-deposits\,\cite{dustgen1} and crack emergence or bifurcation\,\cite{dustgen2}, while the main droplet generation mechanisms are geometry-driven instabilities of shallow melt layers created by edge-localized modes (ELMs), vertical displacement events (VDEs), major disruptions (MDs)\,\cite{dustgen3,dustgen4,dustgen5} and pressure gradient driven melt ejection during unipolar arcing\,\cite{dustgen6,dustgen7,dustgen8}. In this Letter, a novel solid dust generation mechanism is proposed, potentially more harmful than cracking and delamination, since the high speed large solid dust produced can lead to further plasma-facing-component (PFC) damage via subsequent mechanical impacts.

Within the hypervelocity regime, $v_{\mathrm{imp}}\gtrsim3-5\,$km/s, the dust-wall impact speed exceeds the compressional sound speed in both materials generating extreme pressures and temperatures at the collision zone which drive complete dust vaporization and wall crater formation with excavated volumes possibly exceeding those of the dust particle\,\cite{dimpact1}. Within the high-velocity range, $v_{\mathrm{imp}}\gtrsim500\,$m/s, the local pressure build-up remains high enough to severely deform the projectile and to generate a target crater\,\cite{dimpact2}. The first evidence of hypervelocity impacts in FTU was reported more than a decade ago and concerned the in-situ detection of dust impact ionization by electrostatic probes as well as the post-mortem observation of craters on the probe surfaces\,\cite{dimpact3,dimpact4,dimpact5}. However, given the low initial velocities resulting from conventional dust generation mechanisms and dust remobilization\,\cite{dimpact6,dimpact7,dimpact8}, there are no acceleration processes capable of building-up impact speeds above $500\,$m/s. The acceleration problem is fully circumvented in the present mechanism, because the solid dust is generated with very high initial speeds.

\begin{figure}
\centering
\includegraphics[width = 3.3in]{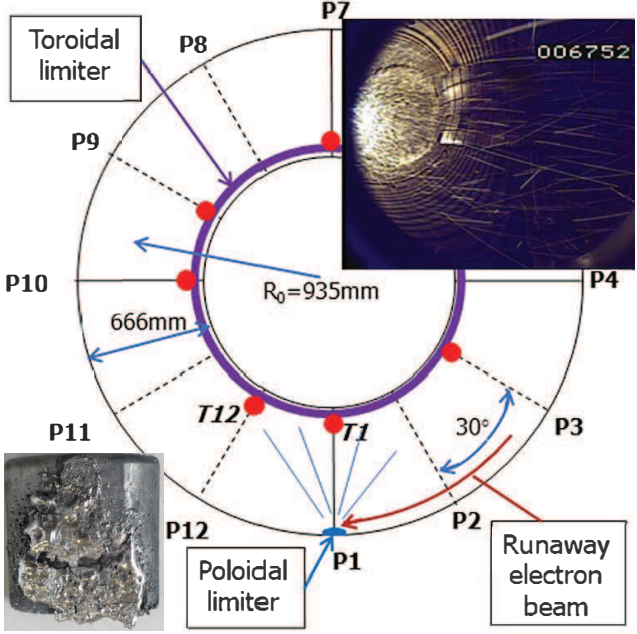}
\caption{Sketch of the equatorial FTU cross-section with the locations of the poloidal and toroidal limiters. Red spots indicate the positions of the tiles observed after the 2013 shut-down. Red arrow indicates the runaway electron trajectories, depending on the $V_{\mathrm{loop}}$ configuration. Top inset: Port P3 VIS camera view of the explosion-like event following the impact of runaway electron beams on the poloidal limiter. Bottom inset: Post-mortem evidence of the explosive runaway electron induced damage at the central tile of the poloidal limiter.}\label{Cross-section}
\end{figure}

\begin{figure}
\centering
\includegraphics[width = 3.1in]{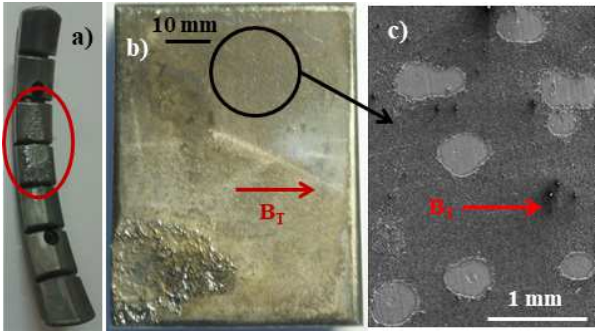}
\caption{(a) Image of the poloidal limiter with the severely molten tiles contained within the red ellipsoid, (b) image of the toroidal tile T12, (c) SEM image of a cratered area detail.}\label{Tiles}
\end{figure}

\emph{Post-mortem observations}. $-$ FTU is a full metal compact tokamak, with toroidal and poloidal limiters made by TZM Mo alloy, that is designed to work with a toroidal magnetic field up to $8\,$T and high plasma densities\,\cite{PizzutoF}. The poloidal limiter is mounted on the equatorial plane of Port P1, at the low field side, while the toroidal limiter is located at the high field side, all around the torus, at the equatorial plane. Visual inspection of the selected limiter tiles, following the 2013 FTU shut-down, revealed severe damage on the equatorial poloidal limiter tile and extensive damage on the toroidal limiter tiles located about in front of Port P1. Tiles from other sectors of the toroidal limiter were unharmed or were far less damaged. Figure \ref{Cross-section} shows the location of these tiles in the vessel. The tiles from the poloidal limiter feature a severely damaged area, see the bottom inset of figure \ref{Cross-section} and figure \ref{Tiles}a, being the preferred damping site for runaway electron beams, while the toroidal tiles T1, T12 feature molten areas along with well pronounced cratered areas, as shown in figure \ref{Tiles}b, c.

The different types of damage observed at different vessel locations can be explained by the following unique sequence of events: runaway electrons impact the poloidal limiter leading to an explosive event that generates large amounts of fast solid dust which impinges on the toroidal limiter tiles leading to extensive crater formation. The proposed mechanism is consistent with the totality of the surface analysis results, presented below.

\begin{figure}
\centering
\includegraphics[width = 3.3in]{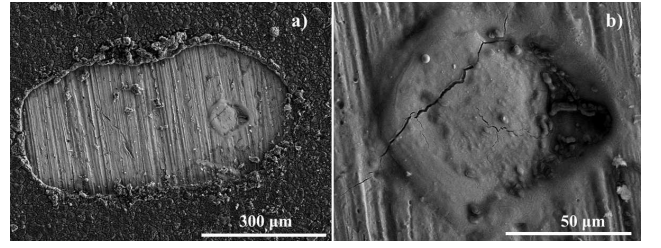}
\caption{(a) SEM image of an elongated halo on tile T12. (b) Zoom-in of the main crater of the same halo.}\label{Crater}
\end{figure}

\emph{Morphological analysis}. $-$ Figure \ref{Crater} illustrates a typical haloed crater, present on tile T12. The tile is covered by a relatively thick co-deposit layer. An extended area can be discerned, wherein no traces of the co-deposit layer are visible and the pristine Mo surface of the tile can be identified by scratches due to the mechanical machining, which corresponds to the \emph{haloed area}. Within each haloed area, a primary hollowed-out area that is located beneath the local surface level can be easily discerned, which corresponds to the \emph{main crater}.

SEM analysis revealed the presence of two types of haloed areas; \emph{near-circular halos} and \emph{elongated halos}, as illustrated in figure  \ref{Tiles}c. The majority of halos detected on tile T12 are elongated and a small fraction is near-circular with the opposite trend observed for tile T1. The correlation of such morphology with the T12-P1 (T1-P1) direction with respect to the T12 (T1) surface normal, see figure \ref{Cross-section}, is apparent. The elongated direction on T12 nearly coincides with the toroidal direction, whereas the elongated halos on T1 are randomly oriented. Extensive SEM and 3D morphological analysis, by means of a mechanical profiler, was carried out which revealed the characteristic dimensions of the haloed areas and the main craters. The results are summarized in Table \ref{morphdim}. Cratering from unipolar arcs can be confidently excluded on the basis of morphological considerations\,\cite{Federici,RohdeV13,Rudako13}.

\begin{figure*}
\centering
\includegraphics[width = 6.2in]{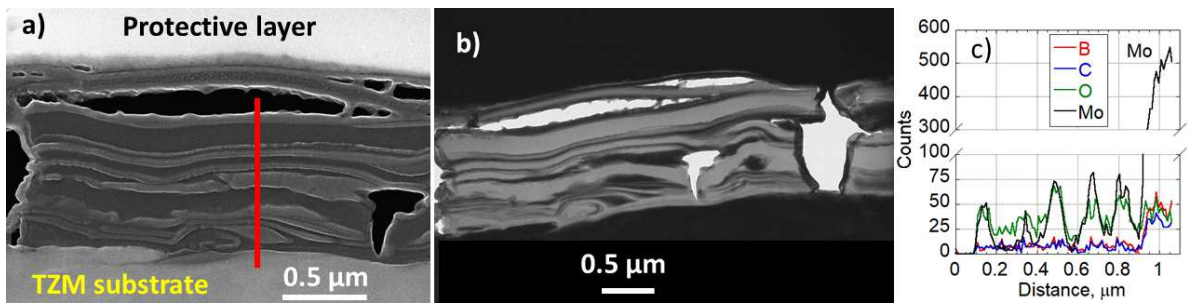}
\caption{Images of a thin foil cut from the co-deposited layer of tile T12 are shown. (a) SE mode image, areas containing heavy elements are brighter; (b) TE mode image, areas rich in light elements are brighter; (c) EDX line scans of Mo, O, B and C across the deposit (red line in figure a, the origin of the distance is at the upper side).}\label{TEM}
\end{figure*}

\emph{Chemical analysis}. $-$ The chemical composition of the co-deposit layer that has grown almost over the entire tile surface can assist in understanding the crater formation mechanism and the type of contaminating elements potentially released upon dust impact. Chemical analysis has been carried out by means of energy-dispersive X-ray spectroscopy (EDX), transmission electron microscopy (TEM), X-ray photoelectron spectroscopy (XPS), secondary ion mass spectrometry (SIMS), and attenuated total reflection Fourier transform infrared spectroscopy (ATR-FTIS). The main results of the analysis for tiles T1 and T12 are: \textbf{(a)} The surface inside the halos is composed of almost pure Mo. \textbf{(b)} The EDX and TEM analyses on the co-deposit layer, see figure \ref{TEM}, revealed the dominant presence of Mo, O, C, and B (due to boronization of the FTU wall). Cr,\,Fe,\,Ni are also present that probably originate from the stainless steel PFCs\,\cite{DeAngeli,SOFT2020,SOFT2021}. \textbf{(c)} XPS and SIMS analyses have confirmed that all metals are present in their metallic and oxidized states and have shown the presence of lithiated compounds in the co-deposit layer\,\cite{Ghezzi18}. \textbf{(d)} ATR analysis confirmed the presence of MoO$_3$ and lithiated compounds such as LiOCH$_3$.

\begin{table}
  \centering
  \caption{The dimensions of the halos and the main craters. $(^1)$ Halo diameter values in case of circular halos.}\label{morphdim}
\begin{tabular}{c c c }
Characteristic        & Circular  & Elongated   \\
Dimension ($\mu$m)    & craters   & craters     \\ \hline\hline
Halo major length$^1$ & $250-400$ & $300-700$   \\
Halo minor length     & N/A       & $200-400$   \\
Main crater diameter  & $50-100$  & $50-90$     \\
Main crater depth     & $3-12$    & $5-10$      \\ \hline\hline
\end{tabular}
\end{table}

Let us consider the worst conditions for co-deposit removal: adopt the maximum observed removed co-deposit volume, assume no layer porosity and no impact-induced flaking, ignore any vaporization prior to the liquid gas transition and assume a $100\%$ MoO$_3$ composition. The minimum energy required to vaporize the co-deposit is
\begin{equation}
E_{\mathrm{v}}=\rho_{\mathrm{co}}V_{\mathrm{co}}\left[c_{\mathrm{p,co}}\left(T_{\mathrm{b}}-T_0\right)+\Delta{H}_{\mathrm{f,co}}+\Delta{H}_{\mathrm{v,co}}\right]\,,
\end{equation}
with $\rho_{\mathrm{co}}$ the co-deposit mass density, $V_{\mathrm{co}}$ the removed volume, $c_{\mathrm{p,co}}$ the co-deposit heat capacity, $T_{\mathrm{b}}=1428\,$K the MoO$_3$ normal boiling point, $T_0=200\,$K the wall temperature, $\Delta{H}_{\mathrm{f,co}}$, $\Delta{H}_{\mathrm{v,co}}$ the latent heats of fusion and vaporization.
The total energy delivered by an incident spinless perfectly spherical dust particle equilibrated with the MoO$_3$ co-deposit up to its boiling point is given by
\begin{equation}
E_{\mathrm{d}}=\frac{1}{6}\pi\rho_{\mathrm{d}}D_{\mathrm{d}}^3\left[\frac{1}{2}v_{\mathrm{imp}}^2+c_{\mathrm{p,d}}\left(T_{\mathrm{d}}-T_{\mathrm{b}}\right)\right]\,,
\end{equation}
with $\rho_{\mathrm{d}}$ the dust mass density, $D_{\mathrm{d}}$ the dust diameter, $v_{\mathrm{imp}}$ the impact velocity, $c_{\mathrm{p,d}}$ the dust heat capacity and $T_{\mathrm{d}}$ the dust temperature. For a $D_{\mathrm{d}}=100\mu$m, $v_{\mathrm{imp}}=1\,$km/s $T_{\mathrm{d}}=2000\,$K pure Mo dust particle, we have $E_{\mathrm{d}}\sim7{E}_{\mathrm{v}}$, which demonstrates that mechanical impacts of fast hot solid dust particles can easily vaporize the co-deposit. Note that the dust thermal energy change (second adder of $E_{\mathrm{d}}$) alone is more than two times larger than $E_{\mathrm{v}}$.

\emph{Explosive dust generation by runaway electron impact}. $-$ Given the morphological and chemical analysis, the only viable explanation is that the craters were generated by the high velocity impact of solid particles of the same material, which were ejected by the poloidal limiter located on the equatorial plane of Port P1. Monitoring video-cameras installed on FTU (25 fps) and an infrared camera (383 fps and $618\,\mu$s of integration time) installed on the Port P1 upper side, \emph{i.e.} just above the poloidal limiter, have acquired multiple images that confirm the rapid ejection of dust from the poloidal limiter. In fact, sequences acquired right after a disruptive event, followed by the generation of runaway electrons (REs) that were localized in a small area at the midplane\,\cite{Ciotti95}, reveal an "explosion-like" event where dust particles are ejected in all directions regardless of the residual plasma current or the magnetic field. The typical values of RE beam energies and currents for FTU are tens of MeV and few hundreds of kA, respectively\,\cite{Esposito}. On the basis of the model described in Ref. \cite{Helander}, estimates were carried out for discharges with RE beams accompanied by such explosion events which led to RE energies of $15-35\,$MeV and RE current lower bounds of $150-230\,$kA, thus confirming the aforementioned expectations. A VIS camera screenshot of an explosive event is provided in the insert of figure \ref{Cross-section}. More important, IR camera images (not shown in figure \ref{Cross-section}) contain particle tracks that are much longer than the $300\,$mm length of the field of view. Given the stated IR integration time, this translates to a $485\,$m/s estimate for the velocity lower bound. The high dust speeds, high poloidal limiter temperatures and large number of particles leaving the field of view in the course of one frame make it impossible to extract the velocity within given uncertainties or to estimate the ejected dust number.

The absence of splashes and the crater morphology necessitate that the dust particles are solid upon impact, but this does not imply that the dust particles are also solid upon generation. In order to elucidate this aspect, dust transport simulations were carried out with the MIGRAINe dust dynamics code\,\cite{MIGRAIN1,MIGRAIN2}. The coupled heating (including phase change) and mass evolution equations were solved for spherical Mo droplets that are free streaming in vacuum. The initial temperature was assumed to be $2900\,$K (barely above the Mo $2896\,$K melting point) and the initial diameter varied within $60-100\,\mu$m. In particular, cooling due to thermionic emission, vaporization and thermal radiation was considered together with mass loss due to vaporization. Owing to the large latent heat of fusion $\sim37.5\,$kJ/mol, full resolidification always required at least $30\,$ms. Assuming a $400\,$m/s speed that lies at the lower side of the high velocity regime, this implies that the droplets traverse at least $12\,$m before resolidifying which is much larger than the distance between the poloidal limiter and tile T12 (see figure\,\ref{Cross-section}). This demonstrates that the fast dust particles must have been generated in the solid phase. Note that any residual plasma (neglected in the simulations) would have further increased the resolidification time and traversed distance.

As aforementioned, off-normal events are known to generate liquid metal droplets rather than solid metallic dust. The fundamental difference between RE and other off-normal heat loads is that REs lead to volumetric heating given the mm depth ranges of $\sim10\,$MeV electrons in PFCs, while ELMs, VDEs or MDs lead to surface heating given the $10-100\,$nm depth ranges of $\sim10\,$keV electrons, protons and helium ions in PFCs\,\cite{MEMOScod}. In addition, the RE energy deposition profile inside PFCs could have a steep maximum that is not located at the surface, since high energy electrons are practically collisionless (the picture becomes more complicated, when electron backscattering, $\delta$- ray generation and bremsstrahlung emission are included)\,\cite{Seltzer&}. Thus, RE-PFC interaction could fall into a particularly challenging regime, where the temperature field attains a rather steep maximum inside the material which opens up the possibility of solid dust ejection during thermal shocks, which are caused by internal thermal stress build-up due to the uneven thermal expansion and the large elastic moduli of refractory metals\,\cite{ThermSh1,ThermSh2,ThermSh3}.

\emph{Crater geometrical reconstruction}. $-$ In order to further support the physics interpretation, the main craters were reproduced by accelerating dust particles towards a stationary bulk target with the aid of a single- or double-stage light gas gun\,\cite{Riva1989,highvel1}. The target was a clean Mo $2\,$mm thickness sheet, the projectiles were spherical Mo dust with diameters in the ranges $35-50\,\mu$m, $50-63\,\mu$m, $63-71\,\mu$m and the controlled impact speeds varied within $200-2200\,$m/s. Tests were carried out at two incident angles with normal impacts emulating crater formation inside circular halos and oblique impacts emulating crater formation inside elongated halos.

\begin{figure}
\centering
\includegraphics[width = 3.3in]{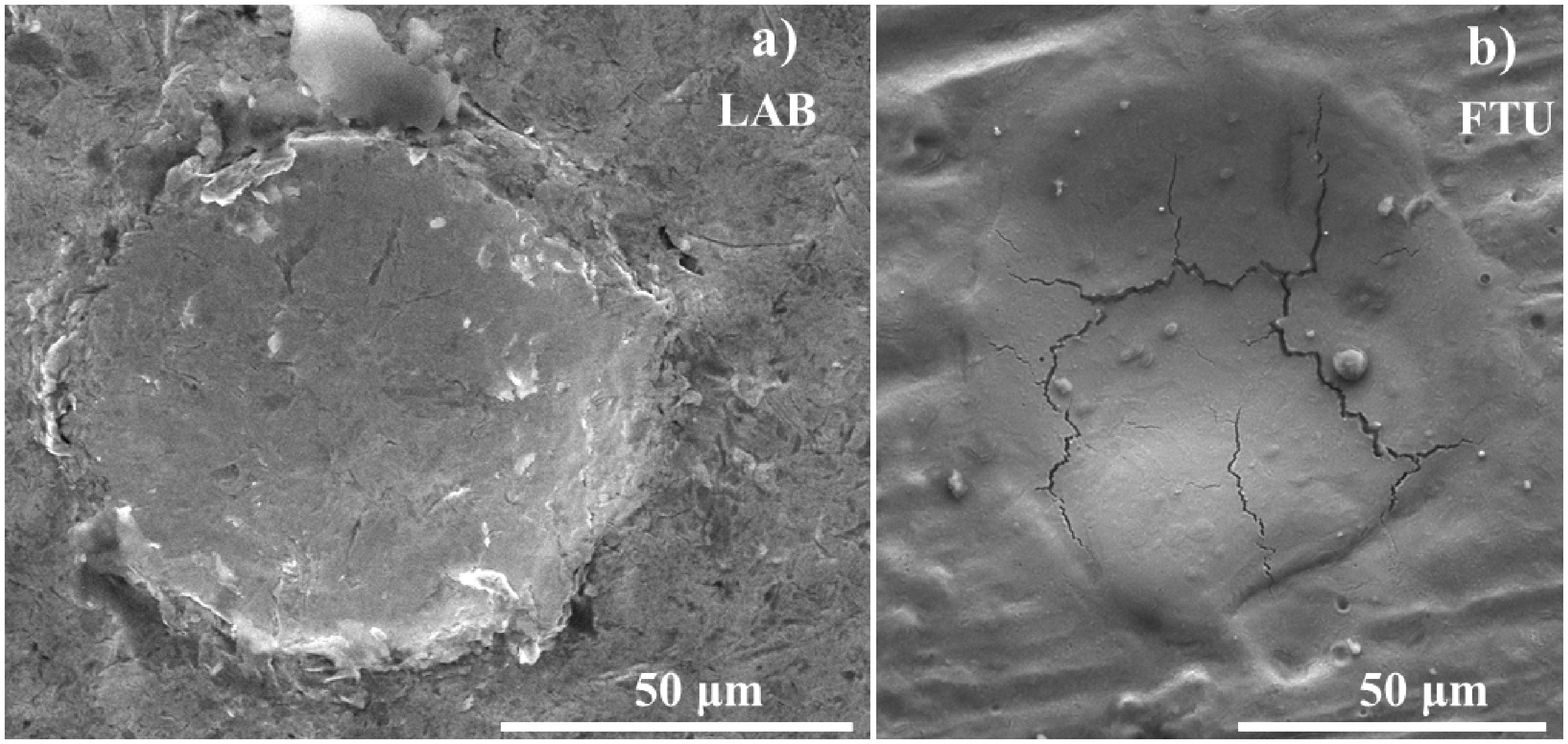}
\caption{Comparison between (a) a crater reconstructed in the laboratory by normal high-velocity dust impact (Mo $63-71\,\mu$m at 688m/s), (b) a main crater inside a near-circular halo of tile T12.}\label{circular}
\end{figure}

\begin{figure}
\centering
\includegraphics[width = 3.3in]{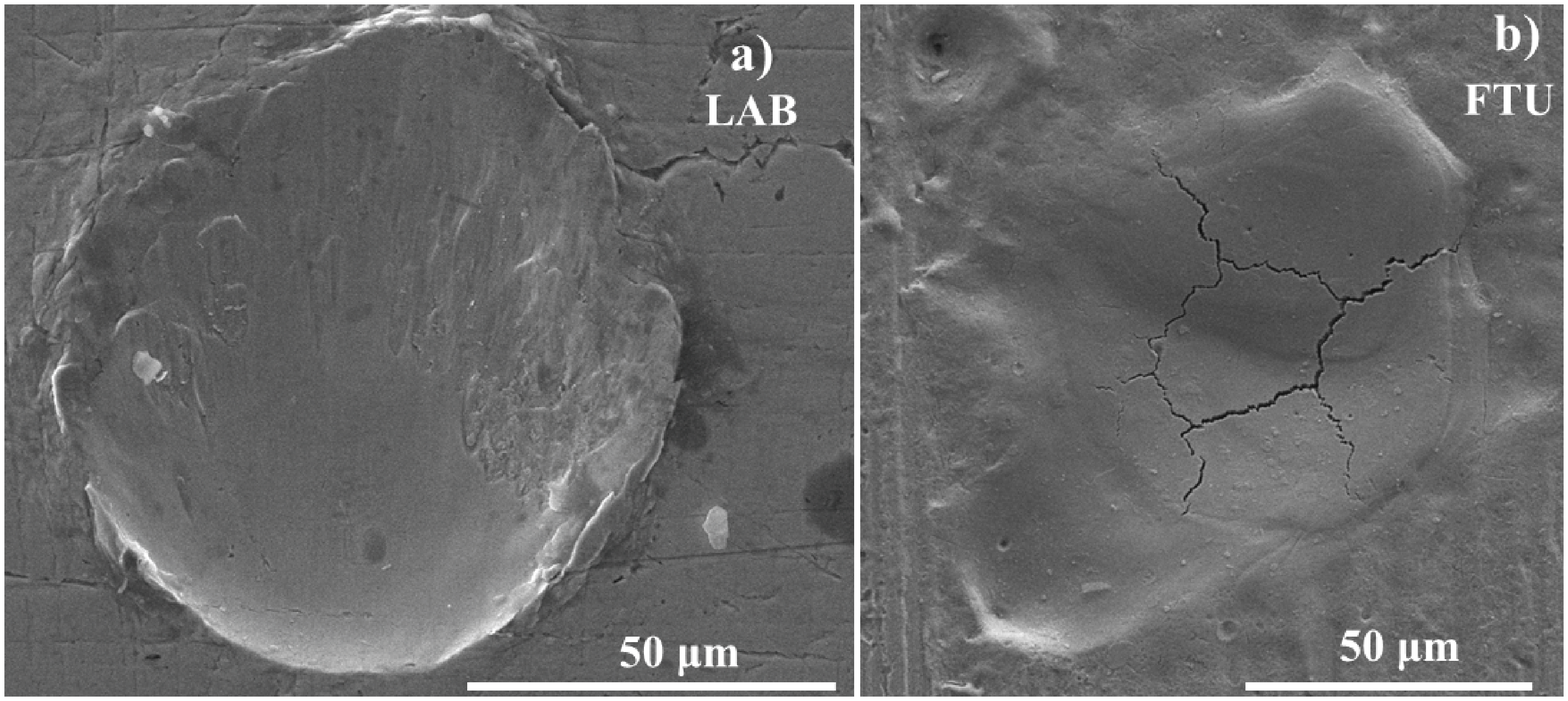}
\caption{Comparison between (a) a crater reconstructed in the laboratory by oblique high-velocity dust impact (Mo $63-71\,\mu$m at 885m/s), (b) a main crater inside an elongated halo of tile T12.}\label{elongated}
\end{figure}

In the case of normal impacts, the crater morphology inside tile T12 was most closely reproduced for $63-71\,\mu$m Mo dust at speeds within $700-800\,$m/s, see also figure \ref{circular}. The typical $77-100\,\mu$m crater diameters and $7-15\,\mu$m crater depths are comparable to the respective dimensions of the craters with circular halos reported in Table \ref{morphdim}. In the case of oblique impacts, the relative position of the poloidal limiter with respect to tile T12  (figure \ref{Cross-section}) suggests an impact angle of $45^{\circ}$ with respect to the normal. The target was tilted accordingly and the crater morphology was best mimicked by $63-71\,\mu$m Mo dust impinging at speeds within $700-900\,$m/s, see also figure \ref{elongated}. The typical $52-80\,\mu$m crater diameters and $3.5-9.5\,\mu$m crater depths are comparable to the respective dimensions of the craters with elongated halos reported in Table \ref{morphdim}.

Naturally, in the laboratory the Mo dust particles and target are at room temperature, while in FTU the TZM dust particles are hot and the TZM tiles are kept between -100 and -70$^{\circ}$C during normal discharges. This translates to morphological crater characteristics that cannot be matched in the laboratory. First of all, FTU craters appear to be re-solidified in contrast to laboratory craters. A first-order energy balance analysis for identical composition projectiles and targets, containing a number of approximations concerning the fraction of the plastic work dissipated as heat, the equipartition of the impact energy between the dust particle and the target, the distribution of the temperature in the affected volume as well as the plastic deformation of the particle, leads to a rather accurate simple semi-empirical expression for the minimum impact speed that triggers melting\,\cite{highvel2}. It reads as
\begin{equation}
v_{\mathrm{imp}}^{\mathrm{melt}}=\left(\frac{20e_{\mathrm{th}}I_{\mathrm{melt}}}{\rho_{\mathrm{d}}\sqrt{D_{\mathrm{d}}}}\right)^{2/5}\,,
\end{equation}
where $e_{\mathrm{th}}=\sqrt{\rho_{\mathrm{d}}k_{\mathrm{d}}c_{\mathrm{p}}}$ is the so-called thermal effusivity with $k_{\mathrm{d}}$ the thermal conductivity and $c_{\mathrm{p}}$ the specific isobaric heat capacity, where $I_{\mathrm{melt}}=T_{\mathrm{m}}-T_0+\Delta{h}_{\mathrm{f}}/c_{\mathrm{pd}}$ is a melting index with $T_{\mathrm{m}}$ the melting point, $T_0$ a characteristic temperature of the pre-impact system and $\Delta{h}_{\mathrm{f}}$ the latent heat of fusion. Its application for $T_0=300\,$K and $D_{\mathrm{d}}=67\,\mu$m yields $v_{\mathrm{imp}}^{\mathrm{melt}}=810\,$m/s, which is consistent with the observation that the $688\,$m/s speed does not generate re-solidified laboratory craters (figure \ref{circular}a). Its application for $T_0=2000\,$K and $D_{\mathrm{d}}=67\,\mu$m yields $v_{\mathrm{imp}}^{\mathrm{melt}}=655\,$m/s, which is consistent with the observation that the $688\,$m/s speed generates re-solidified FTU craters (figure \ref{circular}b). Furthermore, the FTU craters feature cracks in contrast to the reconstructed impact craters (see figure \ref{circular} and \ref{elongated}). The FTU cracks are most likely not a consequence of impact-induced melting followed by rapid re-solidification, since craters from laboratory Mo-on-Mo impacts within $1000-2000\,$m/s, \emph{i.e.} far above the lower threshold melting impact speed, never featured cracks. It is more plausible that the difference in the target temperature could explain this observation, since the TZM ductile-to-brittle transition temperature is $-80^{\circ}\,$C\,\cite{crackin1}. Impact-induced crack formation can be considered as the mechanical equivalent of thermal shock-induced crack formation, which has been consistently observed in ion beam and electron beam high heat flux test facilities\,\cite{crackin2}. Finally, laboratory craters feature a sharp rim in contrast to the FTU craters. The rim is an omnipresent characteristic of Mo-on-Mo but also W-on-W high-velocity impact craters regardless of incident speed\,\cite{crackin3}. However, the rim is not present around Mo-on-W craters at $\sim1000\,$m/s, although it is present around W-on-Mo craters at similar impact speeds. Thus, it is a rather delicate feature that could be attributed to chemical composition mismatches (Mo vs TZM). It should be mentioned that it cannot be excluded that the rim was originally present also in FTU craters, but was eroded away during plasma exposure.

\emph{Comparison with empirical damage laws}. $-$ Systematic hypervelocity and high velocity dust impact experiments aim at the formulation of empirical damage laws for the crater diameter and depth as a function of the material properties, impact speed, impact angle and dust size\,\cite{hyperve1,hyperve2,hyperve3}. Such laws allow for estimates of the total excavated material combined with the additional geometrical assumption of a conical or ellipsoidal crater\,\cite{hyperve4,hyperve5}. Unfortunately, in the high velocity impact regime, most emphasis has been put to the formulation of scaling laws for the critical speeds that define the narrow window for impact-bonding\,\cite{highvel2,highvel3}, \emph{i.e.} the physical phenomenon behind the cold spraying technique. Hence, to benchmark the present laboratory results, established damage laws of the hypervelocity regime would have to be extrapolated down to lower speeds. Since at the high velocity regime the crater diameter is of the order of the dust diameter, we focus on crater depth that has stronger dependence on the impact speed. For metal dust exceeding $50\,\mu$m, the most appropriate damage law reads as\,\cite{hyperve1,hyperve2,hyperve3}
\begin{equation}
D_{\mathrm{c}}=5.24\frac{D_{\mathrm{d}}^{19/18}}{H_{\mathrm{t}}^{1/4}}\left(\frac{\rho_{\mathrm{d}}}{\rho_{\mathrm{t}}}\right)^{1/2}\left(\frac{v_{\mathrm{imp}}\cos{\theta}}{c_{\mathrm{t}}}\right)^{2/3}\,,\label{damagelaw}
\end{equation}
with $D_{\mathrm{c}}$ the crater depth in cm, $D_{\mathrm{d}}$ the dust diameter in cm, $H_{\mathrm{t}}$ the target Brinell hardness, $\rho_{\mathrm{d}},\rho_{\mathrm{t}}$ the dust and target mass densities, $c_{\mathrm{t}}$ the target sound speed, $v_{\mathrm{imp}}$ the dust impact speed and $\theta$ the impact angle with respect to the target surface normal. It is based on impacts for $2\lesssim{v}_{\mathrm{imp}}[\mathrm{km/sec}]\lesssim12$ and $D_{\mathrm{d}}\gtrsim50\,\mu$m. It exhibits excellent agreement with our lab impact data near $2000\,$m/s and satisfactory agreement down to $\sim1000\,$m/s.

\emph{Additional evidence}. In further support of our rationale on the generation of high velocity solid dust by RE impact, we briefly refer to two further independent FTU pieces of evidence. In 2015 and 2016, two bulk Mo targets ($30\times6\mathrm{mm}^2$, $0.5\,$mm thickness, radially directed) were exposed in Port P5 during experiments where REs were often generated. SEM analysis of the molten target area revealed the presence of holes with diameters $50-100\,\mu$m, \emph{i.e.} comparable to the dimensions of the main craters observed on tiles T1, T12 and summarized in Table \ref{morphdim}. Furthermore, after the FTU decommissioning in 2021, the toroidal tiles were collected and the presence of craters, located nearly in the same T12 region, was confirmed. However, it should not be expected that RE interaction with metal PFCs always generates fast hot solid dust. For instance, there has been no such evidence yet in JET with the ITER-Like-Wall\,\cite{dustgen1,dustgen3,dustgen9}.

\emph{Discussion}. Experimental evidence has been presented in favor of a novel solid dust generation mechanism by runway electron explosive incidence. Its most intriguing characteristic concerns the $\sim500\,$m/s initial dust speeds which lead to high-velocity dust-PFC impacts that are accompanied by crater formation and strong dust deformation. Future works will focus on clarifying the runway electron parameters (current density, energy distribution) that are required for explosive PFC damage accompanied by fast dust production as well as on correlating the dust parameters (size and speed distribution) with the runway electron parameters. Finally, it is crucial to investigate the possible occurrence of this novel mechanism in ITER and DEMO, since the explosion-like PFC damage at the primary runaway electron impact site constitutes more extended erosion than conventional melting events and since high-velocity or hypervelocity impacts imply secondary non-localized PFC damage. It is worth mentioning that diffuse interface models provide promising frameworks that could reliably account for runaway electron induced damage and solid dust generation by compressive flows and thermal shock propagation\,\cite{outrodu1,outrodu2}, but no such state-of-the-art computational tool has been yet developed by the fusion community.

\emph{Acknowledgments}. $-$ The authors would like to thank Zana Popovic for valuable discussions about REs in FTU and Monica De Angeli for the acquisition of the profiler images and the analysis of the SEM images. This work has been carried out within the framework of the EUROfusion Consortium, funded by the European Union via the Euratom Research and Training Programme (Grant Agreement No 101052200 — EUROfusion). Views and opinions expressed are however those of the authors only and do not necessarily reflect those of the European Union or the European Commission. Neither the European Union nor the European Commission can be held responsible for them.

\end{document}